\documentclass[11pt]{article}
\usepackage{txfonts}
\usepackage{geometry}
\usepackage[utf8]{inputenc}
\usepackage{enumitem,amssymb}
\usepackage{ragged2e}
\newlist{thematic}{itemize}{8}
\setlist[thematic]{label=$\square$}
\usepackage{pifont}
\usepackage{setspace}
\usepackage{eso-pic,graphicx}
\usepackage{tikz}
\usepackage{atbegshi}
\usepackage{url}
\usepackage{xcolor}    
\usepackage{multicol}
\usepackage[english]{babel}
\usepackage[sort&compress]{natbib}
\usepackage{authblk}
\definecolor{myblue}{RGB}{0,121,194}
\usepackage[colorlinks=true,
            linkcolor=myblue,
            urlcolor=myblue,
            citecolor=myblue]{hyperref}

\begin{document}

\selectlanguage{english}

\raggedright
\Large
ESO Expanding Horizons  \linebreak
\small
Transforming Astronomy in the 2040s \linebreak
Call for White Papers

\vspace{2.cm}
\begin{spacing}{1.6}
\textbf{\fontsize{22pt}{40pt}\selectfont
Beyond Stage IV: Quasar and Galaxy Clustering and the Fundamental Physics of the 2040s}
\end{spacing}
\normalsize
\vspace{0.5cm}

\textbf{Authors:}
M. Guidi$^{1,2}$,
M. Moresco$^{1,2}$,
H. K. Herrera-Alcantar$^{4,5}$,
G. Aricò$^{3}$,
S. Camera$^{16}$, 
C. Carbone$^{6}$,
A. Cimatti$^{1}$,
S. Contarini$^{7}$,
P. Dayal$^{8}$,
G. Degni$^{9}$,
A. Farina$^{10}$,
C. Giocoli$^{2,3}$
V. Iršič$^{15}$, 
A. Labate$^{1,2}$,
F. Marulli$^{1, 2, 3}$,
F. Montano$^{16}$
C. Moretti$^{11}$,
L. Moscardini$^{1}$,
A. Pisani$^{9}$,
A. Pollo$^{12,13}$,
S. J. Rossiter $^{16}$,
E. Sarpa$^{11}$,
S. Sartori$^{9}$,
E. Sefusatti$^{11}$,
M. Talia$^{1,2}$,
F. Verdiani$^{14}$,
A. Veropalumbo$^{10}$ \\
\vspace{0.3cm}

\textbf{Contacts:} \href{mailto:massimo.guidi6@unibo.it}{massimo.guidi6@unibo.it}

\vspace{0.3cm}

\textbf{Affiliations:} \\
$^{1}$ Dipartimento di Fisica e Astronomia, Universit\`a di Bologna, Via Piero Gobetti 93/2, I-40129 Bologna, Italy \\
$^{2}$ INAF--Osservatorio di Astrofisica e Scienza dello Spazio di Bologna, Via Piero Gobetti 93/3, I-40129 Bologna, Italy \\
$^{3}$ INFN--Sezione di Bologna, Viale Berti Pichat 6/2, I-40127 Bologna, Italy \\
$^{4}$ Institut d’Astrophysique de Paris, 98 bis Boulevard Arago, F-75014 Paris, France \\
$^{5}$ IRFU, CEA, Université Paris-Saclay, F-91191 Gif-sur-Yvette, France \\
$^{6}$ INAF--Istituto di Astrofisica Spaziale e Fisica Cosmica di Milano, Via Alfonso Corti 12, I-20133 Milano, Italy \\
$^{7}$ Max Planck Institute for Extraterrestrial Physics, Giessenbachstrasse 1, D-85748 Garching, Germany \\
$^{8}$ Canadian Institute for Theoretical Astrophysics, University of Toronto, Toronto, ON, Canada \\
$^{9}$ Aix Marseille Université, CNRS/IN2P3, CPPM, F-13288 Marseille, France \\
$^{10}$ INAF--Osservatorio Astronomico di Brera, Milano, Italy \\
$^{11}$ INAF--Osservatorio Astronomico di Trieste, Trieste, Italy \\
$^{12}$ National Centre for Nuclear Research, Otwock, Poland \\
$^{13}$ Jagiellonian University, Cracow, Poland \\
$^{14}$ SISSA, Via Bonomea 265, I-34136 Trieste, Italy \\
$^{15}$ Center for Astrophysics Research, Department of Physics, Astronomy and Mathematics, University of Hertfordshire, College Lane, Hatfield AL10 9AB, United Kingdom \\
$^{16}$ Dipartimento di Fisica, Università degli Studi di Torino, via P. Giuria 1, 10125 Torino, Italy
\vspace{0.3cm}

\newpage
\section*{The Challenge: Fundamental Physics Beyond Stage IV}
\justifying

Galaxy and quasar clustering provides unprecedented leverage on the Universe's fundamental laws. Stage IV surveys (DESI, 4MOST, MOONS, Euclid) are delivering transformative baryonic acoustic oscillations (BAO) and redshift space distortions (RSD, \citep{Kaiser1987}) measurements at $z < 2$ using luminous red galaxies and emission-line galaxies, yet several of the most important questions in cosmology will remain inaccessible with current approaches, regardless of sample size increases. Three critical frontiers define the agenda for 2040s:

\textbf{(1) Neutrino Mass Hierarchy:} Despite decades of effort, the absolute neutrino mass scale remains unknown, critical for leptogenesis and beyond Standard Model of Particle Physics \citep{ParticleDataGroup:2024cfk}, and also for the Cosmological Standard Model \citep{Lesgourges2006}. Stage IV surveys, when combined with CMB data, constrain $\Sigma m_\nu \lesssim 0.05\,\mathrm{eV}$ via small-scale clustering suppression, insufficient to resolve the normal vs. inverted mass hierarchy ($\Delta m^2 \sim 0.003\,\mathrm{eV}^2$). Reaching $\Sigma m_\nu \lesssim 0.015\,\mathrm{eV}$ requires: (i) enormous sample volumes to overcome shot noise; (ii) high-redshift coverage ($z > 1.5$) where neutrino effects are maximal; (iii) full-shape clustering analysis ($k$ to quasi-linear scales) where the suppression is sharpest.

\textbf{(2) Multi-field Inflation:} Different multi-field inflationary models predict vastly different primordial non-Gaussianity (PNG)  \citep{Bartolo2005, Donghui2009, Desjacques2010}. Stage IV achieves $\sigma(f_{\mathrm{NL}}) \sim 5$, sufficient to detect PNG if $f_{\mathrm{NL}} \gtrsim 5$, but insufficient to probe the multi-field regime ($f_{\mathrm{NL}} \sim 1$) where different inflation variants make distinctive predictions. The scale-dependent clustering signature of PNG is strongest on the largest observable modes at high redshift, precisely where Stage IV lacks adequate volume and spectroscopic reach to achieve $\sigma(f_{\mathrm{NL}}) \sim 1$.

\textbf{(3) Structure Growth Across Cosmic Time and the Dark Energy Puzzle:} The Universe's expansion and the growth of perturbations encode information about dark energy, gravity, and the dark sector \citep{Peacock2001, Eisenstein2005, Guzzo2008}. Stage IV maps $f\sigma_8(z)$ primarily at $z < 1.5$. However, the diagnostic power lies in the range $1 < z < 3$, where the interplay between ongoing structure growth and emerging dark energy dominance uniquely constrains the nature of dark energy and tests of gravity. Testing gravity and dark sector physics requires: (i) extension to $z > 3$ where deviations from GR are maximal; (ii) dense sampling of $1 < z < 3$ mapping structure growth evolution precisely as dark energy dominated; (iii) full-shape clustering accessing both expansion history and small-scale physics where screening mechanisms and quantum corrections to gravity may leave observable imprints. Together, these measurements trace the Universe's structural evolution across half its age, revealing the nature of dark energy and whether gravity requires modification or new dark sector physics.

\section*{What Stage IV Cannot Deliver}

Stage IV surveys face fundamental limitations:
\begin{itemize}
\item \textbf{Redshift Coverage:} DESI reaches $z \sim 2$ for ELGs; 4MOST/MOONS optimized for $z \lesssim 1.5$. At $z > 2$, only sparse quasar populations exist \citep{Shen2007}. Yet this is where neutrino suppression, PNG amplification, and gravity tests are most constraining.

\item \textbf{Spectroscopic Precision:} Current DESI redshift 
accuracy is $\sigma_z \sim 0.00003(1+z)$. Full-shape clustering and optimal neutrino mass constraints benefit from redshift precisions approaching $\sigma_z \lesssim 0.00001(1+z)$, a 3$\times$ 
improvement demanding revolutionary spectroscopic and calibration capabilities.

\item \textbf{Multi-Tracer Synergy:} Stage IV uses multiple tracers independently. Joint analysis across continuous redshifts and luminosities, enabling self-calibrating bias and systematic control, is not achievable with current survey designs.
\end{itemize}

\section*{A Transformative Vision}

Transcending Stage IV requires a qualitative paradigm shift:

\begin{description}
\item[Dense High-Redshift Sampling:] A survey targeting $1 < z < 6$ with uniform density across multiple tracer populations (star-forming galaxies, emission-line galaxies, LRGs, quasars). This demands revolutionary multi-object spectroscopy: simultaneous observation of $\mathcal{O}(1000)$ targets per pointing with consistent, high-resolution spectra.

\item[Large Uniform Sky Coverage:] $\geq 10{,}000\,\mathrm{deg}^2$ with consistent depth to access cosmologically representative volumes and sample the largest modes where PNG and gravity signatures dominate.

\item[Precision Spectroscopy at Scale:] Achieving $\sigma_z/(1+z) \lesssim 0.0002$ across tens of millions of objects via advanced calibration, stable multi-year operations, and rigorous systematic error control.
\end{description}

\section*{Science Payoff: Breakthrough Constraints}

Such a survey would deliver:
\begin{itemize}
\item \textbf{Neutrino Physics:} $\Sigma m_\nu \lesssim 0.015\,\mathrm{eV}$ (3$\times$ Stage IV improvement), providing cosmological constraints complementary to neutrinoless double beta decay searches, relevant to the Majorana versus Dirac nature of neutrinos.

\item \textbf{Primordial Physics:} Detection or constraints on $f_{\mathrm{NL}}^{\mathrm{local}} \sim 1$, directly probing multi-field inflation and beyond-Standard-Model early-Universe physics.

\item \textbf{Structure Growth Mapping:} $f\!\sigma_8(z)$ measurements across $1 < z < 6$, testing modified gravity and dark energy evolution over half the Universe's age.

\item \textbf{Precision Cosmology:} Percent-level constraints on $D_A(z)$, $H(z)$ at multiple redshifts, breaking degeneracies and testing spatial curvature.
\end{itemize}

These are not incremental gains but qualitative leaps, enabling discovery-driven cosmology rather than precision refinement of a concordance model.

\section*{Technology Requirements}

\begin{description}
\item[Ultra-High-Multiplexing Spectroscopy:] Flexible fiber or IFU systems with $\mathcal{O}(1000)$ simultaneous spectra per pointing, enabling dense sampling across diverse source populations and redshifts.

\item[Spectroscopic Precision:] Wavelength calibration and systematic control at the $\sigma_z \lesssim 0.0002(1+z)$ level sustained over multi-year campaigns.

\item[Data Infrastructure:] Advanced pipelines for redshift determination, full-shape modeling across multiple tracers, and rigorous systematic error budgeting.
\end{description}

\section*{The Cosmological Frontier: Toward Discovery Beyond Precision}

The 2040s mark a transformative inflection point for cosmology. Stage IV surveys establish a precision foundation; in the near future a qualitatively new capability is needed to unlock the physics of neutrinos, inflation, and gravity that will remain fundamentally inaccessible otherwise. 

A next-generation spectroscopic survey delivering dense, high-redshift, full-shape clustering across vast sky areas represents more than an incremental advance: it enables a paradigm shift in how cosmological information is extracted. Beyond traditional two-point statistics (BAO, RSD), higher-order clustering statistics encode unique signatures of primordial non-Gaussianity, neutrino masses, and gravity itself, breaking many of the degeneracies that characterize two-point statistics alone. Together with field-level inference techniques that exploit the full information content of three-dimensional matter distributions and advanced machine learning methods for optimal summary statistics and systematic error characterization, such a survey would unlock information currently hidden in the data noise.

This convergence of unprecedented data volume, multi-tracer synergies, and transformative analysis techniques, spanning from large-scale structure to quasi-linear scales and incorporating higher-order statistics and field-level inference, would fundamentally advance our ability to constrain the deepest open questions in cosmology. Precision measurements of neutrino masses, constraints on inflationary physics through primordial non-Gaussianity, and rigorous tests of gravitational physics across cosmic epochs would transition from speculative to experimentally determined. 

Concept studies for wide-field spectroscopic facilities in the 2040s already outline many of the required capabilities. In particular, designs featuring a very wide field of view, high- and low-resolution multi-object spectroscopy, and integral-field spectroscopy over large areas point to a concrete technical route toward such a survey. Recent studies of next-generation wide-field spectroscopic telescopes for ESO, including white papers on the proposed Wide-Field Spectroscopic Telescope (WST) concept \citep[e.g.][]{WST_WhitePaper}, demonstrate that combining these elements into a single facility is both technically feasible and scientifically compelling. The necessary observational and analytical infrastructure to address these questions is therefore not speculative, but a realistic prospect for the 2040s and beyond.
\small
\bibliographystyle{unsrt}
\bibliography{bibliography}
\end{document}